\documentclass[final,3p,times]{elsarticle}
               
\usepackage{graphicx}
\usepackage{amssymb}
\usepackage{amsmath}
\usepackage{hyperref} 
\usepackage{longtable}
\biboptions{sort&compress}
\usepackage{xcolor}

\newcommand{\transpose}{\intercal}

\journal{Annals of Physics}

\newcommand{\preprint}{
 \setlength{\unitlength}{1mm}{\hbox{\begin{picture}(0,0)
      \put(160,10){\mbox{\footnotesize%
        ADP-13-20/T840}}\end{picture}}}}

\begin{document}

\begin{frontmatter}

\title{\preprint  
       Searching for low-lying multi-particle thresholds in lattice spectroscopy}

\author[adl,csiro]{M. Selim Mahbub}
\author[adl]{Waseem Kamleh}
\author[adl]{Derek B. Leinweber}
\ead{derek.leinweber@adelaide.edu.au}
\author[adl]{Anthony G. Williams}
\address[adl]{Special Research Centre for the Subatomic Structure of
  Matter, School of Chemistry and Physics, The University of Adelaide, SA, 5005, Australia.}
\address[csiro]{CSIRO Computational Informatics, College Road, Sandy
  Bay, TAS 7005, Australia.}

\begin{abstract}
We explore the Euclidean-time tails of odd-parity nucleon correlation
functions in a search for the $S$-wave pion-nucleon scattering-state
threshold contribution.  The analysis is performed using $2+1$ flavor
$32^{3}\times 64$ PACS-CS gauge configurations available via the ILDG.
Correlation matrices composed with various levels of fermion
source/sink smearing are used to project low-lying states.  The
consideration of 25,600 fermion propagators reveals the presence of
more than one state in what would normally be regarded as an
eigenstate-projected correlation function.
This observation is in accord with the scenario where the eigenstates
contain a strong mixing of single and multi-particle states but only
the single particle component has a strong coupling to the
interpolating field.
Employing a two-exponential fit to the eigenvector-projected
correlation function, we are able to confirm the presence of two
eigenstates.  The lower-lying eigenstate is consistent with a $N \pi$
scattering threshold and has a relatively small coupling to the
three-quark interpolating field.
We discuss the impact of this small scattering-state contamination in
the eigenvector projected correlation function on previous results
presented in the literature.
\end{abstract}


\end{frontmatter}

\section{Introduction}

The hadron spectrum provides an interesting foundational platform with
which to investigate the QCD interactions of quarks and gluons.  It
presents significant challenges to current investigations of this
relativistic quantum field theory.  How do the resonances observed in
experiment emerge from the first principles of QCD?  What is the
structure of these states and can it be linked to known effective
degrees of freedom?  For example, are elusive states like the
$\Lambda(1405)$ or the nucleon Roper resonance exotic, perhaps having
a molecular meson-baryon structure?

In this paper we address the first question by performing a Lattice
QCD study of the nucleon spectrum in a search for the multi-particle
scattering threshold states which ultimately generate the finite width
of the resonances in the infinite volume limit.
Correlation matrices composed of traditional three-quark operators
have been very successful in revealing a dense spectrum of baryon
excited states in lattice QCD
\cite{Bulava:2009jb,Mahbub:2010rm,Bulava:2010yg,Engel:2010my,%
Menadue:2011pd,Edwards:2011jj,Mahbub:2012ri,Lang:2012db,%
Mahbub:2013ala,Engel:2013ig,Alexandrou:2013fsu,Morningstar:2013bda}.
However the lowest lying multi-particle scattering state thresholds
are often absent in the observed spectra.

The coupling of these two-particle dominated states to localized
three-quark operators is suppressed relative to single-particle
dominated states.  In full QCD, 3-quark operators will have some
coupling to the meson-baryon components of QCD eigenstates through
interactions with the sea-quark loops of the QCD vacuum.  However,
this coupling is small relative to the coupling to the
single-particle three-quark component of the eigenstate.  

When the three-quark operator creates a resonance in the infinite
volume limit, the overlap with a state dominated by a meson-baryon
component is suppressed on the finite lattice volume, $V$, as
$V^{-1/2}$.  On large volumes these multi-particle dominated states
will be difficult to observe with three-quark operators alone.

In the large-volume case, it is the mixing of one- two- and multi-particle
components in the finite-volume QCD eigenstates that predominantly
governs the presence of multi-particle states when using traditional
three-quark operators alone.  As discussed in detail in the following,
these multi-particle threshold scattering states are likely hidden
within the projected correlation functions of correlation matrices
composed purely of three-quark interpolating fields.  Our focus here is
to reveal these low-lying hidden states.

In the following, we report a case where two states are indeed
participating in what otherwise would be considered to be an
eigenstate-projected correlation function.  Through a two-state
analysis of the projected correlator we are able to accommodate this
weakly coupled second state and evaluate the extent to which it
influences the determination of the mass of the dominant state.

\section{Correlation Matrix Techniques}
\label{CorrelationMatrixTechniques}

To isolate energy eigenstates we use the correlation matrix or
variational method~\cite{Michael:1985ne,Luscher:1990ck}.  To access
$N$ states of the spectrum, one requires a minimum of $N$
interpolators.  With the assumption that only $N$ states contribute
significantly to the correlation matrix $G_{ij}$ at time $t$, the
parity-projected two-point correlation function matrix for $\vec{p}
=0$ can be written as
\begin{align}
G_{ij}^{\pm}(t) &= \sum_{\vec x}\, {\rm Tr}_{\rm sp}\, \{
\Gamma_{\pm}\, \langle\, \Omega\, \vert\, \chi_{i}(x)\,
\bar\chi_{j}(0)\, \vert\, \Omega\, \rangle\}, 
\label{eq:CM} \\
          &=\sum_{\alpha}^N\, \lambda_{i}^{\alpha}\,
\bar\lambda_{j}^{\alpha}\, e^{-m_{\alpha}t},
\end{align}
where Dirac indices are implicit, $\lambda_{i}^{\alpha}$ and
$\bar\lambda_{j}^{\alpha}$ are the couplings of interpolators
$\chi_{i}$ and $\bar\chi_{j}$ at the sink and source respectively,
$\alpha$ enumerates the energy eigenstates with mass $m_{\alpha},$ and
$\Gamma_{\pm}=(\gamma_{0}\pm 1)/2$ projects the parity of the
eigenstates.  

Using an average of $\{U\} + \{U^*\}$ configurations, our construction
of $G_{ij}^{\pm}(t)$ is symmetric and real.  We enforce this symmetry
by working with the improved unbiased symmetric construction $(G_{ij}
+ G_{ji})/2$.
To ensure that the matrix elements are all $\sim{\cal{O}}(1)$, each
element of $G_{ij}(t)$ is normalized \cite{Mahbub:2013ala} by the
diagonal elements of $G(0)$ as ${G}_{ij}(t) / (
  \sqrt{{G}_{ii}(0)}\, \sqrt{{G}_{jj}(0)} )$ (no sum on $i$ or $j$).

An operator creating state $\alpha$ can be constructed as
$\bar{\phi}^{\alpha}=\sum_{j}{\bar\chi}_{j}\, u_{j}^{\alpha}$.  As the
time dependence of the two-point function is governed by $\exp(
-m_\alpha\, t)$ a recurrence relation can be used to solve for
$u_{j}^{\alpha}$
\begin{align}
G_{ij}(t_{0}+\triangle t)\, u_{j}^{\alpha} & = e^{-m_{\alpha}\triangle
  t}\, G_{ij}(t_{0})\, u_{j}^{\alpha}  \, .
 \label{eq:recurrence_relation}
\end{align}  
Multiplying from the left by $G^{-1}(t_0)$ provides the right
eigenvector equation for $u_{j}^{\alpha}$
\begin{equation}
[(G(t_{0}))^{-1}\, G(t_{0}+\triangle t)]_{ij}\, u^{\alpha}_{j} =
c^{\alpha}\, u^{\alpha}_{i} \, , 
\label{eq:right_evalue_eq}
\end{equation} 
with $c^{\alpha}=e^{-m_{\alpha}\triangle t}$.  Similarly, an operator
annihilating state $\alpha$ can be defined as $\phi^\alpha =\sum_j
\chi_j\, v_j^\alpha$, where $v_j^\alpha$ is given by the left
eigenvalue equation
\begin{equation}
v^{\alpha}_{i}\, [G(t_{0}+\triangle t)\, (G(t_{0}))^{-1}]_{ij} =
c^{\alpha}v^{\alpha}_{j} \, .
\label{eq:left_evalue_eq}
\end{equation} 
The eigenvectors for state $\alpha$, $u_{j}^{\alpha}$ and
$v_{i}^{\alpha}$, provide the eigenstate projected correlation
function
\begin{align}
G^{\alpha}_{\pm}(t) & \equiv v_{i}^{\alpha}\, G^{\pm}_{ij}(t)\, u_{j}^{\alpha} ,
\label{projected_cf_final}
\end{align}
with parity $\pm\,$.  We note that with our symmetric construction for 
$G_{ij}^{\pm}(t)$, the left and right eigenvectors are equal.

A eigenvector analysis of a symmetric matrix having orthogonal
eigenvectors can be constructed by inserting
${G^{-1/2}(t_{0})}\, {G^{1/2}(t_{0})}=I$ in
Eq.~(\ref{eq:right_evalue_eq}) and multiplying by
${G^{1/2}(t_{0})}$ from the left,
\begin{align}
{G^{-1/2}(t_{0})}\, G(t_{0}+\triangle t)\, {G^{-1/2}(t_{0})}\,
{G^{ 1/2}(t_{0})}\, u^{\alpha} & = c^{\alpha}\,
{G^{ 1/2}(t_{0})}\, u^{\alpha}\, ,
\label{eqn:symmetric_evalue_deriv} \\
{G^{-1/2}(t_{0})}\, G(t_{0}+\triangle t)\, {G^{-1/2}(t_{0})}\,
w^{\alpha} & = c^{\alpha}\, w^{\alpha} \, ,
\label{eqn:symmetric_evalue}
\end{align} 
where, $w^\alpha = {G^{1/2}(t_{0})}\, u^\alpha$ and
$[{G^{-1/2}(t_{0})}\, G(t_{0}+\triangle t)\,
  {G^{-1/2}(t_{0})}]$ is a real symmetric matrix, with
orthogonal eigenvectors ${w}^{\alpha}$. 

We normalize the eigenvectors ${w}^{\alpha\transpose} {w}^{\beta} =
\delta^{\alpha\beta}$ and define
\begin{align}
u^\alpha = G^{-1/2}(t_0)\, w^\alpha \, ,
\label{eqn:u_def}
\end{align}
and similarly for $v^\alpha$, such that the projected correlator
\begin{align}
 G^{\alpha}(t) & \equiv v^{\alpha\transpose }\, G(t)\, u^{\alpha} \nonumber \\
 & = w^{\alpha\transpose }\, G^{-1/2}(t_0)\, G(t)\, G^{-1/2}(t_0)\, w^\alpha \, ,
\label{projected_cf_final_w}
\end{align}
equals 1 at $t=t_0$.  This construction holds the advantage of
correlating the uncertainties relative to the correlation function at
variational parameter time $t_0$.

In constructing the correlation matrix we consider the local nucleon
interpolating fields $\chi_A = \epsilon^{abc} (u^{a\transpose}\, C
\gamma_5\, d^b)\, u^{c}$ and $\chi_B =
\epsilon^{abc}(u^{a\transpose}\, C \, d^b)\, \gamma_5\, u^{c}$,
commonly referred to as $\chi_1$ and $\chi_2$ in the literature.
Gauge-invariant Gaussian smearing~\cite{Gusken:1989qx} is used to
enlarge the basis of operators. Four different smearing levels are
used at the fermion source and sink for each of the two nucleon
interpolators, providing an $8 \times 8$ basis.

\section{Multi-particle State Contributions}

When using traditional three-quark operators alone in constructing the
correlation matrix on a large volume lattice, it is the mixing of
one- two- and multi-particle components in QCD eigenstates that
predominantly governs the presence of multi-particle components in the
finite-volume eigenstates.

To better understand this mechanism, consider for example the
following simple two-component toy model of two QCD energy
eigenstates, $|\, a \rangle$ and $|\, b \rangle$.  Consider the case
where each state is composed of a localized single-hadron component
denoted by $|\, 1 \rangle$, and a meson-baryon component denoted by
$|\, 2 \rangle$ with arbitrary mixing governed by $\theta$
\begin{eqnarray}
\big|\, a \big\rangle &=& \;\;\, \cos \theta\, \big|\, 1 \big\rangle + \sin \theta\: \big|\, 2 \big\rangle\, , \\
\big|\, b \big\rangle &=& -      \sin \theta\, \big|\, 1 \big\rangle + \cos \theta\, \big|\, 2 \big\rangle\, , 
\end{eqnarray}
Suppose our three-quark interpolator (which may be a linear
superposition of three-quark interpolators from the correlation matrix
analysis) only has significant coupling with $|\, 1 \rangle$. That is
\begin{equation}
\big\langle \Omega\, \big |\, \phi^a_{3q} \, \big |\, 1 \rangle \propto Z \, ,\quad \mbox{and} \quad
\big\langle \Omega\, \big |\, \phi^a_{3q} \, \big |\, 2 \rangle \ll Z \, .
\end{equation}
In this case, acting on the QCD vacuum with $\overline \phi^{\,a}_{3q}$ will
create a superposition of QCD eigenstates as
\begin{equation}
\big|\, 1 \big\rangle = \cos \theta\, \big|\, a \big\rangle - \sin \theta\: \big|\, b \big\rangle \, ,
\end{equation}
and the two-particle components will appear in each of the QCD
eigenstates as they are resolved through Euclidean time evolution.  In
the absence of an operator sensitive to the $|\, 2 \rangle$ component
of the states, it is not possible to disentangle the two QCD energy
eigenstates in the projected correlator.  The projected correlator
contains a superposition of the two states.  A similar discussion can
be made for isospin-1 $\pi \pi$ $P$-wave scattering contributions to
the vector meson correlator.~\cite{Dudek:2012xn}.

Consider further the specific case where the mixing angle $\theta$ is
not too large such that state $|\, a \rangle$ is dominated by a single
particle component and state $|\, b \rangle$ is predominantly a
meson-baryon state.  If we further set their masses $M_a > M_b$ then
we are describing the scenario where the resonance like state $|\, a
\rangle$ dominates the lattice correlation function but a small
admixture of state $|\, b \rangle$ also participates in the lattice
correlation function through the mixing of one and two-particle
components in the QCD eigenstates.  In the absence of an interpolating
field having substantial overlap with $|\, 2 \rangle$ the projected
lattice two-point correlation function will always be composed of the
two QCD energy eigenstates as
\begin{eqnarray}
G_{2pt}(t) &\sim& Z^2\, \cos^2 \theta\, \exp\left(-M_a\, (t-t_s)\right
) \nonumber \\
           &+&    Z^2\, \sin^2 \theta\, \exp\left(-M_b\, (t-t_s)\right )
                \, ,
\end{eqnarray}
and for sufficiently large Euclidean time, $t$, relative to the source
time $t_s$, the lower-lying state will be revealed in the tail of the
lattice correlation function.

However, when $M_a$ and $M_b$ do not differ significantly there is a
concern that the presence of the second state will not be observed in
a $\chi^2/{\rm dof}$ analysis.  Instead, its undetected presence will
change the slope of $\log G_{2pt}(t)$ and thus alter the determination
of mass $M_a$.

\section{Simulation Techniques}

We use the PACS-CS $2+1$ flavor dynamical-fermion
configurations~\cite{Aoki:2008sm} made available through the ILDG
\cite{Beckett:2009cb}.  These configurations use the
non-perturbatively ${\cal{O}}(a)$-improved Wilson fermion action and
the Iwasaki-gauge action~\cite{Iwasaki:1983ck}.  The lattice volume is
$32^{3}\times 64$, with $\beta=1.90$ providing a lattice spacing
$a=0.0907$ fm with the physical lattice volume of $\approx (2.90\,
\rm{fm})^{3}$.

The degenerate up and down quark masses are considered with the
hopping parameter value of $\kappa_{ud}= 0.13770$ and the strange
quark $\kappa_{s}=0.13640$ providing a pion mass of $m_{\pi}$ = 0.293
GeV~\cite{Aoki:2008sm}.  We consider four fermion sources on each of
400 gauge field configurations equally spaced in the time direction.
Configurations are circularly shifted in the time direction after
which a fixed boundary condition is introduced at $t=N_t=64$.  The
fermion source is placed away from the boundary at $t_s=N_t/4=16$ such
that hadron masses extracted from the large Euclidean time tails of
the correlators are maximally displaced from the boundary.
Gauge-invariant Gaussian smearing~\cite{Gusken:1989qx} is used at the
fermion source and sink with a fixed smearing fraction and four
different smearing levels including 16, 35, 100 and 200
sweeps~\cite{Mahbub:2010rm,Menadue:2011pd}.  This provides a total of
25,600 fermion propagators in the correlation matrix analysis.

Our selection of a fixed boundary condition prevents states from
wrapping around the lattice and enables one to carefully examine the
exponential time dependence without significant artifacts.  Only the
pion correlator lives long enough to reveal the effect of the fixed
boundary condition in our simulations.  As the lowest mass hadron with
the longest correlation length, the pion correlator provides the most
stringent test for boundary effects.  From $t=49$, the pion effective
mass systematically rises more than one standard deviation above the
normal fluctuations observed.  We note that this is 15 time slices
from the boundary at 64.

The ground-state nucleon correlator does not survive long enough to
see the boundary with the uncertainty exceeding the signal at $t=42$.
No systematic drift is observed in the correlator prior to signal
loss.  Similarly, the signal in the odd-parity correlator of interest,
examined in detail in the following, is lost at $t=30$.  As this is 34
time slices from the boundary, the determination of the properties of
the low-lying scattering state observed herein is well displaced from
the boundary.

The effective mass function is defined as
\begin{equation}
M_{\rm eff}(t) = \frac{1}{n} \log\left ( \frac{G(t)}{G(t+n)} \right )
\, .
\end{equation}
In presenting our results we will refer to effective mass functions
generated with $n=1$ or 2, noting that $n=2$ provides greater control
in the evaluation of the mass at the expense of reducing the number of
points illustrated before the correlator is lost to noise.

As described in detail in the following section, a second-order
single-elimination jackknife analysis \cite{Efron:1979eb} provides the
uncertainties with the ${\chi^{2}}/{\rm{dof}}$ obtained via the full
covariance matrix analysis.

\section{Jackknife Error Analysis}

Let us consider a single Monte-Carlo sample for one of the matrix
elements of $G_{ij}^{\pm}(t)$ of Eq.~(\ref{eq:CM}) and refer to this
sample as $C_k(t)$ where the subscript identifies the $k$'th
configuration of $N_{\rm con}$ configurations considered in
constructing the ensemble average or mean
\begin{equation}
\overline C(t) = \frac{1}{N_{\rm con}} \sum_{k=1}^{N_{\rm con}} C_k(t)
\, .
\label{eq:mean}
\end{equation}
To simplify the following discussion, we will suppress the time
dependence of $C$ noting the relations below are to be applied to
each time slice.

Because a single Monte-Carlo sample, $C_k$, is not
necessarily\footnote{ A classic example is estimating $\pi$ by
  counting the number of randomly distributed points within the $1
  \times 1$ square falling within $x^2 + y^2 = 1$.  While a single
  Monte-Carlo sample is either 1 or 0 (inside or outside the arc) an
  average over many samples estimates $\pi/4$.} an approximation to
the ensemble average, $\overline C$, it is essential to only consider
averaged quantities when estimating uncertainties.
To this end, the single-elimination jackknife sub-ensemble is
introduced \cite{Efron:1979eb}
\begin{eqnarray}
\overline C_i &=& \frac{1}{N_{\rm con} - 1} \sum_{k=1 \atop k \ne
  i}^{N_{\rm con}} C_k \, , \\
&=& \frac{N_{\rm con}\, \overline C - C_i}{N_{\rm con} - 1} \, ,
\label{eq:jse}
\end{eqnarray}
representing the ensemble average without consideration of the $i$'th
configuration. 
Defining the average of the jackknife sub-ensembles in the usual manner
\begin{equation}
\overline{\overline C} = \frac{1}{N_{\rm con}} \sum_{k=1}^{N_{\rm
    con}} \overline C_k \, ,
\label{eq:jseAvg}
\end{equation}
the standard deviation of the mean, 
$\sigma_C$, is given by 
\begin{equation}
\sigma_C^2 = \frac{N_{\rm con} - 1}{N_{\rm con}} 
\sum_{k=1}^{N_{\rm con}}
\left ( \overline C_k - \overline{\overline C} \right )^2 \, .
\label{eq:sigma}
\end{equation}

We note that in the case where a single $C_k$ {\it is} an
approximation of $\overline C$, Eqs.~(\ref{eq:jse}) and
(\ref{eq:jseAvg}) can be used to take Eq.~(\ref{eq:sigma}) to the
familiar form
\begin{equation}
\sigma_C^2 = \frac{1}{N_{\rm con}} \frac{1}{N_{\rm con} - 1} 
\sum_{k=1}^{N_{\rm con}}
\left ( C_k - {\overline C} \right )^2 \, .
\label{eq:sigmaNaive}
\end{equation}
The change in the leading factor by $(N_{\rm con} - 1)^2$ reflects the
fact that $\overline C_k$ is $(N_{\rm con} - 1)$ times more accurate
than a single $C_k$ and its presence in the square on the right-hand
side of Eq.~(\ref{eq:sigma}).

Turning our attention to the time dependence of $\overline C(t)$ we
note that fluctuations in $\overline C_i(t)$ and $\overline C_i(t+n)$ for
small values of $n=1,\ 2,\ 3,\ldots$ are correlated as these
time slices are next to each other on the lattice and the importance
sampling of the lattice action establishes relationships between the
time slices.  In evaluating the fit of $\overline C(t)$ to a
theoretical model $T(t)$ over a range of time slices
from $t_0$ through $t_f$ one must take these correlations into
account.

The covariance matrix is a generalisation of Eq.~(\ref{eq:sigma}) that
allows for this correlation to be included
\begin{eqnarray}
V(t_i, t_j) &=& \frac{N_{\rm con} - 1}{N_{\rm con}} 
\sum_{k=1}^{N_{\rm con}}
\left ( \overline C_k(t_i) - \overline{\overline C}(t_i) \right )
\left ( \overline C_k(t_j) - \overline{\overline C}(t_j) \right )
 \, , \label{eq:covMat} \\
&=& \left ( N_{\rm con} - 1 \right ) \left [
\frac{1}{N_{\rm con}} 
\sum_{k=1}^{N_{\rm con}}
\overline C_k(t_i) \, \overline C_k(t_j) - 
\overline{\overline C}(t_i) \, \overline{\overline C}(t_j) 
\right ] \, .
\end{eqnarray}
If $\overline C(t_i)$ is not correlated with $\overline C(t_j)$ for
$t_i \ne t_j$, then $V(t_i, t_j)$ becomes diagonal with $V(t_i, t_i) =
\sigma_C^2(t_i)$.

With the jackknife estimate of the covariance matrix, the full
$\chi^2$ including correlations in the data can be evaluated 
\begin{equation}
\chi^2 = \sum_{t_i, t_j} 
\left ( \overline C(t_i) - T(t_i) \right ) \,
C^{-1}(t_i, t_j) \,
\left ( \overline C(t_j) - T(t_j) \right ) \, ,
\label{eq:covChi2}
\end{equation}
where $t_i$ and $t_j$ take all time values $t_0$ through $t_f$
addressing all elements of the inverse covariance matrix, $C^{-1}(t_i,
t_j)$.  The inverse is calculated via the singular value decomposition
algorithm.  In counting the associated degrees of freedom for the
$\chi^2$, one counts the number of time slices considered in the fit,
$N_t$, and reduces by the number of parameters in the theoretical
model and the number of singular values encountered in inverting
$C(t_i, t_j)$.  

In the case where $C^{-1}(t_i, t_j)$ is diagonal, $C^{-1}(t_i, t_i) =
1 / \sigma_C^2(t_i)$ and Eq.~(\ref{eq:covChi2}) provides the familiar
measure
\begin{equation}
\chi^2 = \sum_{t_i} \frac{ 
\left ( \overline C(t_i) - T(t_i) \right )^2
}{
\sigma_C^2(t_i)} \, .
\label{eq:chi2}
\end{equation}
However, Eq.~(\ref{eq:chi2}) will substantially underestimate the
$\chi^2$ if the data are correlated, as the matrix sum of $N_t^2$
values of $t_i$ and $t_j$ has been reduced to the $N_t$ diagonal
entries $t_i = t_j$.  Thus, the full covariance-matrix based
$\chi^2/{\rm dof}$ is required to evaluate the fit and all
$\chi^2/{\rm dof}$ quoted in this study are from the full covariance
matrix.

While the presentation to this point is sufficient to determine the
uncertainties on $G_{ij}^{\pm}(t)$ of Eq.~(\ref{eq:CM}) and enable a
fit, one also desires uncertainties on the fit parameters.  A
second-order single-elimination jackknife provides these.  One
proceeds by defining a jackknife sub-ensemble in which two different
configurations have been removed from the average
\begin{eqnarray}
\overline C_{ij} &=& \frac{1}{N_{\rm con} - 2} \sum_{k=1 \atop k \ne
  i,\ k \ne j}^{N_{\rm con}} C_k \, , \\
&=& \frac{N_{\rm con}\, \overline C - C_i - C_j}{N_{\rm con} - 2} \, ,
\label{eq:jseSecOrd}
\end{eqnarray}
representing the ensemble average without consideration of the $i$'th
nor the $j$'th configurations.  The index $j$ of $\overline C_{ij}$
can be ``jackknifed'' to get the uncertainty for the correlator
$\overline C_i(t),$
\begin{equation}
\sigma_{C_i}^2 = \frac{N_{\rm con} - 2}{N_{\rm con} - 1} 
\sum_{j=1 \atop j \ne i}^{N_{\rm con}}
\left ( \overline C_{ij} - \overline{\overline C}_i \right )^2 \, ,
\label{eq:sigmaSecOrd}
\end{equation}
where
\begin{equation}
\overline{\overline C}_i = \frac{1}{N_{\rm con} - 1} \sum_{k=1 \atop k
\ne i}^{N_{\rm con}} \overline C_{ik} \, ,
\label{eq:jseAvg2ndOrd}
\end{equation}
defines the average of the second-order jackknife sub-ensembles.  The
covariance matrix is given by the generalisation of
Eq.~(\ref{eq:sigmaSecOrd}) where the squared factor at a single time
is replaced by the same factor at two different times.

A fit to $\overline C_i \pm \sigma_{C_i}$ produces a fit parameter
such as the baryon mass, $M_i$.  The uncertainty for $M$ from a fit to
the ensemble average can be obtained by ``jackknifing'' the $i$ index
of $M_i$ via Eq.~(\ref{eq:sigma}) with $C_i \to M_i$
\begin{equation}
\sigma_M^2 = \frac{N_{\rm con} - 1}{N_{\rm con}} 
\sum_{i=1}^{N_{\rm con}}
\left ( \overline M_i - \overline{\overline M} \right )^2 \, .
\label{eq:sigmaM}
\end{equation}

In our calculations, all quantities are combined at the same order of
jackknife such that the error analysis takes into account all
correlations and the final error estimates provide an accurate
estimate of the statistical uncertainty.

\section{Results}

Here we focus on the odd-parity sector, $G_{-}^\alpha(t)$, seeking
evidence of the low-lying $N \pi$ $S$-wave scattering threshold
state.  This threshold state is notably absent in most lattice QCD
calculations and will reveal itself in the large Euclidean-time tail
of the correlation function.

In fitting the projected correlation function, we seek a fit composed
of a minimum of four points in $G_{-}^\alpha(t)$.  
The lower and upper time limits of the fit window are denoted by
$t_{\rm{min}}$ and $t_{\rm{max}}$ respectively.
We commence by setting $t_{\rm {min}}$ equal to the lead variational
time parameter $t_{0}$, and $t_{\rm{max}}$ to the last time slice with
the uncertainty in $G_{-}^\alpha(t)$, $\triangle G_{-}^\alpha(t) <
G_{-}^\alpha(t)$.  Occasionally the correlation function displays a
transition to noise and a lower value of $t_{\rm{max}}$ is set.  An
example of this is provided in the following.
The ${\chi^{2}}/{\rm{dof}}$ is limited to $\le 1.30$ as larger values
usually introduce a systematic error in the extracted mass.  In
searching for a satisfactory fit we first reduce $t_{\rm{max}}$ and
only increase $t_{\rm {min}}$ if an acceptable fit providing a
$\chi^2/{\rm dof} < 1.3$ is not obtained.  We do not place a lower
limit on the $\chi^2/{\rm dof}$ as small values typically reflect
large uncertainties as opposed to an incorrect result associated with
a systematic error.

If there are exactly $N$ states contributing in a significant manner
to an $N\times N$ correlation matrix analysis and the basis of the
correlation matrix spans the eigenstate space, then a fit commencing
at $t_{\rm {min}} = t_{0}$ should be possible.

\begin{figure}[t]
  \begin{center} 
  \includegraphics[height=0.46\textwidth,angle=90]{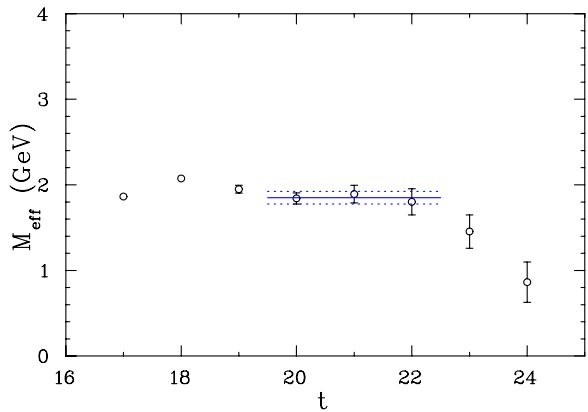}
 \end{center}
\vspace{-0.5cm}
 \caption{(Color online). Effective mass function for the second
   $N{1/2}^{-}$ state over 350 configurations from
   Ref.~\cite{Mahbub:2012ri}.  The best fit with a mass of 1.85(7) GeV
   and $\chi^{2}/{\rm{dof}}=0.50$, is also shown.}
 \label{fig:meff_n-_350cfg}
\end{figure}

We refer to the states observed in our correlation matrix analysis as
the first, second, third, $\ldots$, odd-parity states, with the first
state being the lowest energy state.  Remarkably, the correlation
function for the first state shows no evidence of a lower-lying
scattering threshold state in the Euclidean-time tail.  Therefore we
turn our attention to the second state of our correlation matrix
analysis.  We note that this state has relatively strong overlap with
our $\chi_B$ interpolating field, in contrast to the first state which
is dominated by $\chi_A$.

Figure~\ref{fig:meff_n-_350cfg} presents the effective mass, $M_{\rm
  eff}(t) = \log( G(t)/G(t+1))$, from Ref.~\cite{Mahbub:2012ri} for
the second odd-parity state from 350 configurations.  The correlation
matrix analysis was performed at $t_0 = 18$ relative to the source at
$t_s=16$ with $\triangle t = 2$ such that one seeks a fit commencing
at $t_{\rm {min}} = t_{0} = 18$.  $t_{\rm{max}}= 23$ was initialized
as the effective mass at $t=24$ was viewed as a transition to noise
which commences at $t=25$.  The fit satisfying our criteria is
illustrated commencing at $t_{\rm {min}} = 20$, two time steps after
$t_{0}$.

There are two scenarios that could lead to $t_{\rm {min}} > t_{0}$.
In the first and most familiar scenario, the number of states
participating in the correlation functions of the $8 \times 8$
correlation matrix exceed 8 and the higher-energy state contaminations
are introducing curvature at early times.  In this case further
Euclidean time evolution is required to reduce the contributions of
the highest states in the spectrum to an insignificant level such that
only 8 states are significant in the correlation matrix analysis.
Further discussion of this issue is included in the Appendix of
Ref.~\cite{Mahbub:2013ala}.

In practice, one can implement the correlation matrix analysis at
later variational times.  However, uncertainties grow rapidly.  We
have noted an insensitivity of the eigenvectors to the variational
parameters.  This is reflected in the fact that the extracted masses
are consistent and insensitive to the variational parameters of
$t_{0}$ and $\triangle t$.  In conclusion, we accept $t_{\rm{min}}$ in
the range $t_0 \le t_{\rm{min}} \le t_{0} + \triangle t$ as providing
the best estimate of the eigenstate energy.

In this scenario, where high-energy states are inducing curvature in
the correlation function at early times, a lower-lying $N \pi$
scattering state may be present in the projected correlation function.
However, its contribution is suppressed relative to the dominant state
and its presence results in a negligible systematic error.

In the second scenario, the contribution from a lower lying $N \pi$
scattering state in the projected correlation function is significant.
The combination of two states gives rise to curvature in the effective
mass at $t = 18$ and 19 following the source at $t_s=16$.  The small
$\chi^{2}/{\rm{dof}}=0.50$ of the fit in Fig.~\ref{fig:meff_n-_350cfg}
provides no hint of a second state and the extracted mass represents a
superposition of two states as opposed to a finite-volume QCD
eigenstate.  In this case the reported mass will contain an undetected
systematic error.

The presence and strength of a lower lying $N \pi$ scattering state
will be revealed in the large Euclidean time tail of the correlation
function.  Thus to explore these two scenarios further, we quadruple
the number of fermion sources on each configuration and use the full
set of 400 configurations available from PACS-CS via the ILDG.  In the
figures, we refer these results as `1600 cfg' and contrast these
results with the earlier `350 cfg' results \cite{Mahbub:2012ri}.

\begin{figure}[tb!]
  \begin{center} 
  \includegraphics[height=0.46\textwidth,angle=90]{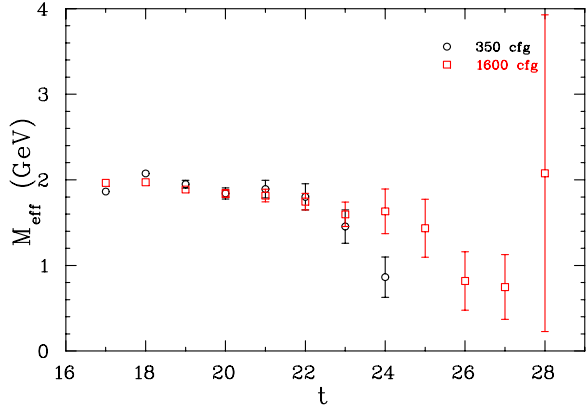}
  \end{center}
\vspace{-0.5cm}
 \caption{(Color online). Effective mass function $M_{\rm eff}(t) =
   (1/2) \log( G(t)/G(t+2))$ for the second $N{1/2}^{-}$ state for
   1600 configurations is compared with previous results from 350
   configurations~\cite{Mahbub:2012ri}.  A fit from $t=20$ to 27
   provides a covariance-matrix based $\chi^{2}/{\rm{dof}}=2.82$ and
   rejects the hypothesis of a single state at a 99.4\% confidence
   level.  The presence of a lower-lying state is manifest in the tail
   of the effective mass.}
 \label{fig:meff_n-_350cfg_1600cfg_state2.1By2lngtBygtPlus2}
\end{figure}

Figure~\ref{fig:meff_n-_350cfg_1600cfg_state2.1By2lngtBygtPlus2}
illustrates the effective mass obtained from 1600 fermion sources of four
different smearing extents at the source and sink; i.e. 25,600 quark
propagators.  
The presence of a second lower-lying contribution to this projected
correlation function is now manifest in the drift of the effective
mass as a function of Euclidean time.  A fit from $t=20$ to 27
provides a covariance-matrix based $\chi^{2}/{\rm{dof}}=2.82$.  With
seven degrees of freedom the $\chi^{2}$ distribution rejects the
hypothesis of a single state at the 99\% confidence level and instead
indicates the presence of an additional state(s).  The new results
also confirm that the 350 configuration result at $t=24$ is in fact
due to a loss of signal near the onset of noise at $t=25$.

We note that the rejection of the single-state hypothesis contrasts
studies of the isovector vector-meson channel of the $\rho$ meson
\cite{Dudek:2012xn} where no evidence of two-particle $\pi \pi$
scattering contributions to the $\rho$-meson correlator was observed
when using two-quark operators alone.  Only with the specific
introduction of four-quark operators, could the $\pi \pi$
contributions to the channel be resolved.  This is in accord with the
very different nature of the quark flow diagrams and associated
couplings describing meson dressings of mesons and baryons in QCD
\cite{Allton:2005fb,Armour:2008ke}.

Having confirmed the presence of at least two states in the projected
correlation function, we now consider two-state fits to the projected
correlation function.

As explained in Sec.~\ref{CorrelationMatrixTechniques}, our
normalization of the orthogonal eigenvectors $w^\alpha$ and subsequent
definitions of $u^\alpha$ and $v^\alpha$ provide $G^\alpha(t_{0}) = 1$
for the projected correlation function.
This constraint reduces the standard two-exponential fit function with
four parameters 
\begin{equation}
G^\alpha(t) = \lambda_{1}\, \exp \left ( -M_{1}\, (t-t_{s}) \right )
            + \lambda_{2}\, \exp \left ( -M_{2}\, (t-t_{s}) \right ) \, ,
\end{equation}
to a three parameter function with 
\begin{equation}
\lambda_{2} = \frac{ 1 - \lambda_{1} \exp \left ( -M_{1}\, (t_{0}-t_{s}) \right ) }
                   { \exp \left ( -M_{2}\, (t_{0}-t_{s}) \right )} \, .
\end{equation}
This construction ensures $G^\alpha(t_{0}) = 1$ exactly and thus the
$\chi^2/{\rm dof}$ is evaluated over the interval $t=t_{0}+1$ to
$t_{\rm{max}}$.  As
Fig.~\ref{fig:meff_n-_350cfg_1600cfg_state2.1By2lngtBygtPlus2}
indicates a loss of signal in $G^\alpha(t)$ at $t = 28 + 2 = 30$, we
commence with the largest interval having $t_{\rm{max}} = 29$.

\begin{figure*}[t!]
  \begin{center} 
\includegraphics[height=0.48\textwidth,angle=90]{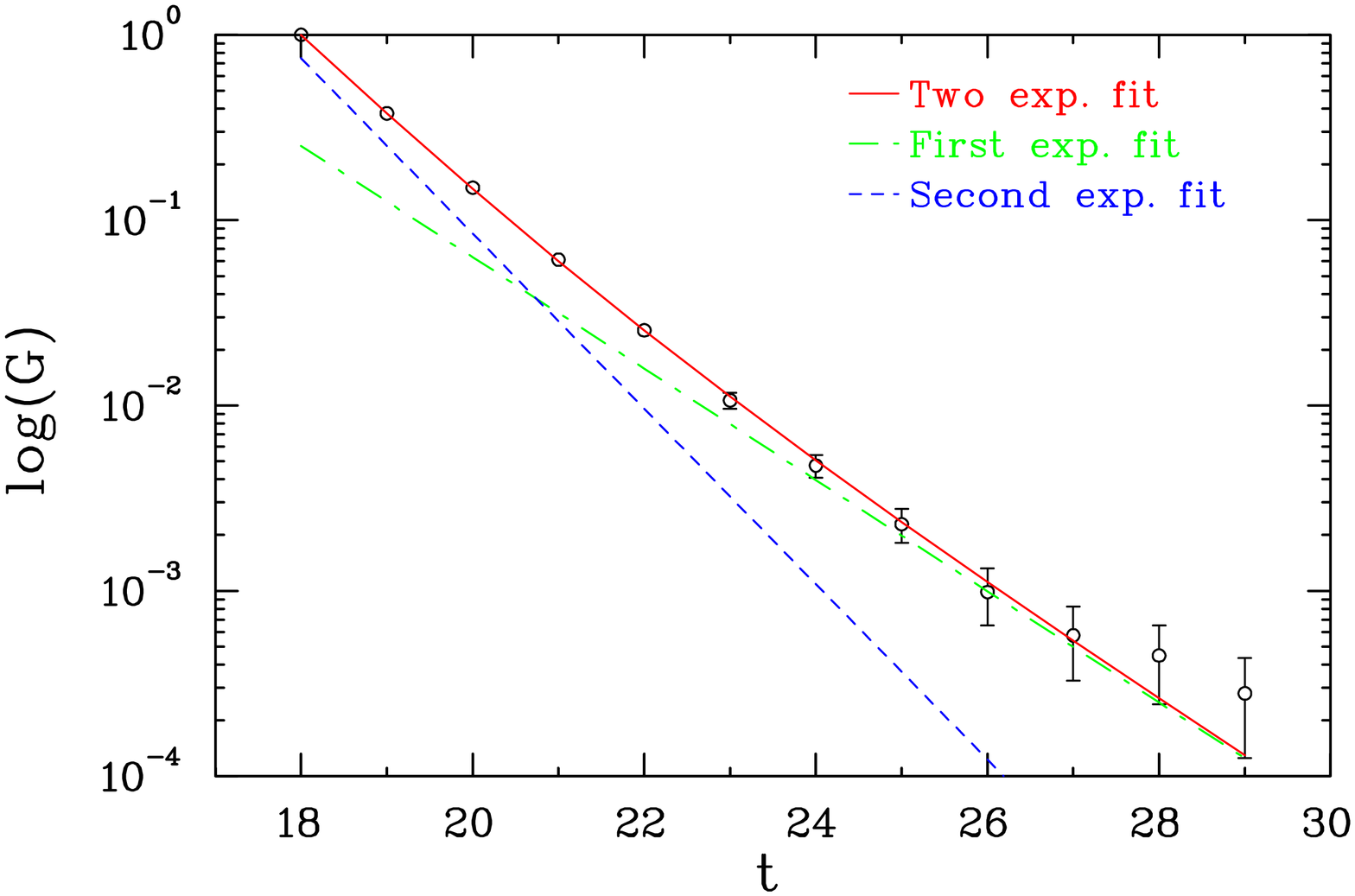}
\quad
\includegraphics[height=0.48\textwidth,angle=90]{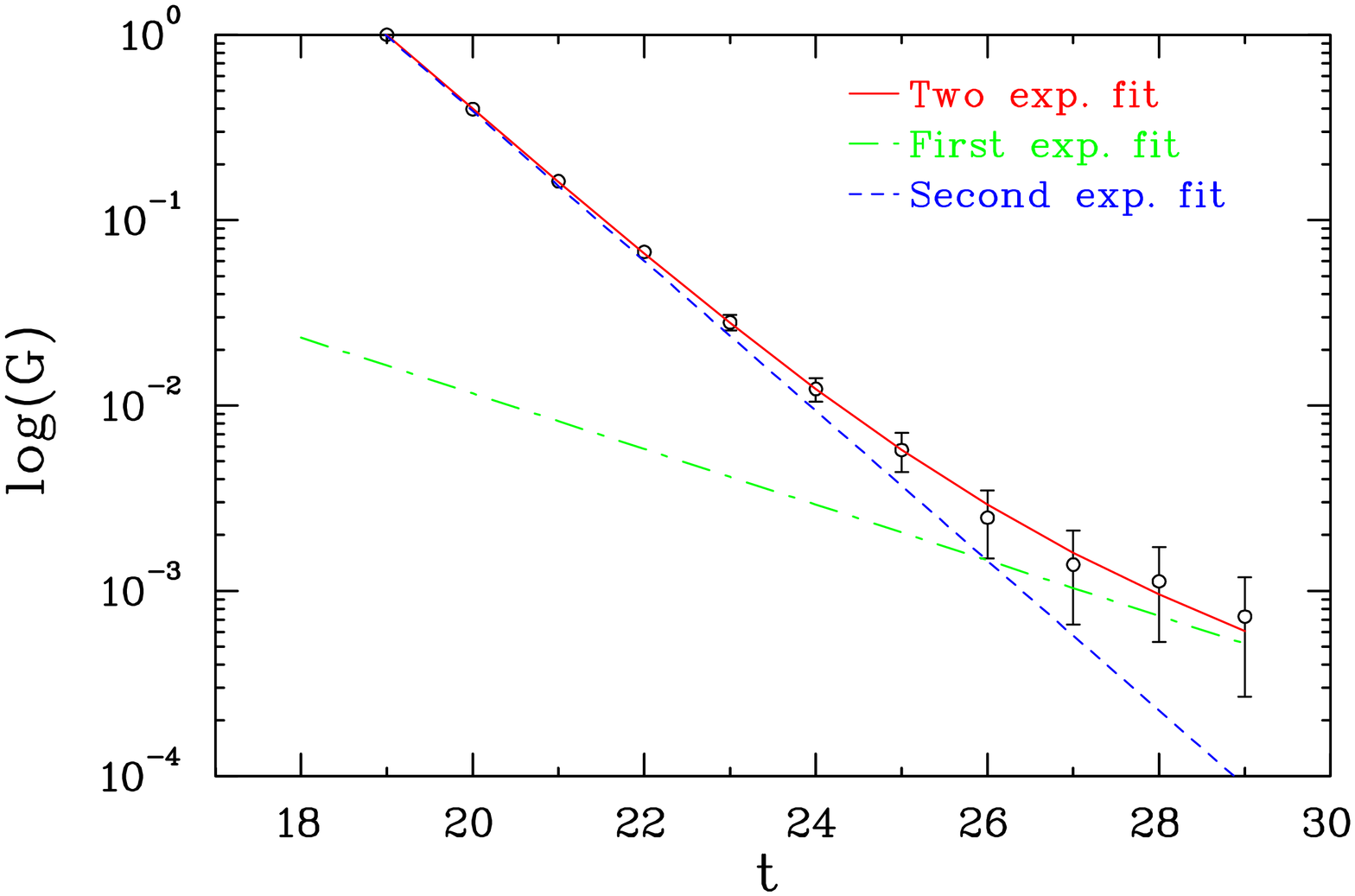} 
 \end{center}
\vspace{-0.6cm}
 \caption{(Color online). (Left) A two exponential fit to the
   projected correlation function obtained from a generalized
   eigenvalue analysis at $t_0 = 18$, $t_0 + \triangle t = 20$
   relative to the fermion source at $t_s=16$.  (Right) A similar fit
   to the projected correlation function obtained from a generalized
   eigenvalue analysis at $t_0 = 19$, $t_0 + \triangle t = 20$.
   Dashed and dash-dot lines illustrate the individual exponentials
   from the fits while the full line presents the sum of exponentials
   fit to the lattice results.
}
 \label{fig:data_and_fit}
\vspace{-0.2cm}
\end{figure*}

Figure~\ref{fig:data_and_fit} illustrates two-state fits to the
projected correlation function of the second state of the correlation
matrix spectrum.  The left-hand plot displays the results of a
correlation matrix analysis at $t_0=18$, $t_0 + \triangle t = 20$
relative to the fermion source at $t_s=16$.  The right-hand plot
illustrates results for a correlation matrix analysis at $t_0=19$,
$t_0 + \triangle t = 20$.  In both cases, the $\chi^{2}/{\rm{dof}}$ is
well below one.  As is common for two-state fits, the nature of the
fits depends sensitively on the earliest time slice considered in the
fit.

\subsection{$t_0 = 18$ Analysis}

Including results at $t=18$ and 19 in the fit, prior to the onset of
the plateau in
Fig.~\ref{fig:meff_n-_350cfg_1600cfg_state2.1By2lngtBygtPlus2} at
$t=20$ results in a fit where both states are given similar weight.
The main role of the additional state is to accommodate the curvature
in $\log(G)$ at early times.  This is consistent with the second
scenario described earlier in this section.

Setting $t_0$ one step later leads to a very different fit where the
additional state is given a very small weight and its main role is to
accommodate curvature in the tail of the correlation function.  This
is consistent with the first scenario described earlier herein.

Table \ref{table:two_exp_fit} presents variational parameters, fit
results, correlated ratios and the $\chi^{2}/{\rm{dof}}$ for these two
fits as well as several other closely related fits.  The large
uncertainties in the results illustrate the interplay between the two
exponentials and the importance of establishing correlation matrices
that are able to couple strongly to the two-particle components of the
QCD eigenstates and enable the isolation of each state.

\begin{table*}[tb!]
   \begin{center}
   \vspace{-0.4cm}
   \caption{Fitted parameters including masses (in GeV) and coupling
     strengths ($\lambda$) from two-exponential fits to projected
     correlation functions obtained with variational parameters
     $t_{0}$ and $\triangle t$ in an $8\times 8$ correlation matrix
     analysis.  Fits are from $t_0$ to $t_{\rm max}$ relative to the
     source at $t_s=16$.  The ratio of $M_1$ to $M_2$ and
     $\lambda_{1}$ to $\lambda_{2}$ and their correlated errors are
     also shown.
     Note, the infinite volume scattering threshold is $M_N + m_\pi =
     1.36$ GeV at this second lightest quark mass of the PACS-CS
     configurations and is expected to be attractive on the finite
     volume of the lattice.
}
   \vspace{0.2cm}
   \label{table:two_exp_fit}
   \begin{tabular}{ccccccccccc} 
    \hline
\noalign{\vspace{3pt}}
    $t_{0}$ & $\triangle t$ &$t_{\rm max}$ & $M_{1}$ & $M_{2}$ & $M_1/M_2$ &
    $\lambda_{1}$ & $\lambda_{2}$ & $\lambda_1/\lambda_2$ & $\chi^{2}/{\rm{dof}}$ \\
\noalign{\vspace{3pt}}
    \hline 
    \hline
\noalign{\vspace{3pt}}
    18 & 1 & 28  & 1.54(25)  & 2.45(41) & 0.62(03) & 1.83(1.95) & 6.22(1.23) & 0.29(37)  & 0.50 \\  
    18 & 2 & 28  & 1.53(39)  & 2.36(50) & 0.65(05) & 1.60(2.83) & 6.19(2.02) & 0.26(56)  & 0.48 \\  
    18 & 3 & 28  & 1.56(43)  & 2.37(60) & 0.65(05) & 1.75(3.38) & 6.02(2.48) & 0.29(71)  & 0.48 \\  
    18 & 1 & 29  & 1.49(30)  & 2.38(40) & 0.62(03) & 1.48(2.02) & 6.43(1.28) & 0.23(36)  & 0.47 \\  
    18 & 2 & 29  & 1.43(49)  & 2.26(41) & 0.63(11) & 1.00(2.53) & 6.60(1.77) & 0.15(44)  & 0.36 \\  
    18 & 3 & 29  & 1.45(56)  & 2.25(49) & 0.64(12) & 1.05(3.04) & 6.52(2.20) & 0.16(56)  & 0.35 \\  
    19 & 1 & 28  & 0.91(85)  & 1.95(11) & 0.46(41) & 0.12(0.77) &16.25(0.97) & 0.01(05)  & 0.11 \\  
    19 & 2 & 28  & 1.06(99)  & 1.97(20) & 0.53(48) & 0.25(2.54) &16.31(1.58) & 0.01(16)  & 0.16 \\  
    19 & 1 & 29  & 0.71(68)  & 1.93(06) & 0.37(34) & 0.04(0.20) &16.05(0.92) & 0.002(12) & 0.10 \\   
    19 & 2 & 29  & 0.78(85)  & 1.93(08) & 0.41(43) & 0.06(0.40) &16.09(1.03) & 0.004(24) & 0.10 \\  
\noalign{\vspace{3pt}}
   \hline 
 \end{tabular}
 \end{center}
\end{table*}

While the selection of $t_0$ governing where the fit starts plays a
significant role, the variation of $\triangle t$ has a negligible
effect on the results.

When commencing at $t_0 = 18$, reducing $t_{\rm{max}}$ by one to 28
has little effect on the results.  Here the focus is on small times
where the uncertainties are small.  $M_1 = 1.54(25)$ GeV compares
favorably with the infinite volume scattering threshold of $M_{N} +
m_{\pi} = 1.35$ GeV at this second lightest quark mass of the PACS-CS
configurations.  However, one is anticipating an attractive
interaction on the finite volume lattice and in this light the lattice
value is somewhat large.  

The best determined value for $M_2 = 2.4(4)$ GeV is larger than the
published result \cite{Mahbub:2012ri} of 1.85(7) GeV illustrated in
Fig.~\ref{fig:meff_n-_350cfg} and presents an explicit case of how an
undetected scattering state could contribute to the slope of the lattice
correlation function and mask the true mass of the dominant state.

While each of the masses are not accurately determined, the mass ratio
is, due to the strong correlation between these two parameters.
Similarly the amplitudes of the states are poorly determined but the
ratio of the amplitudes $\lambda_1/\lambda_2$ is the order of $1/10$,
large enough to provide an important systematic error as described
earlier.

\subsection{$t_0 = 19$ Analysis}

Turning our attention to $t_0 = 19$, the inclusion of $t_{\rm{max}} =
29$ is of some assistance, better constraining both masses.  The most
accurate result for $M_2$ of 1.93(6) GeV compares favorably with the
published result from 350 configurations \cite{Mahbub:2012ri} of
1.85(7) GeV.  It also agrees well with the same fit of the effective
mass from 1,600 configurations producing 1.84(5) GeV, with the
$\chi^{2}/{\rm{dof}}= 0.3$.  This small $\chi^{2}/{\rm{dof}}$ for a
single-state fit provides further support that the right-hand panel of
Fig.~\ref{fig:data_and_fit} with $t_0$ conservatively delayed to $t =
19$ is the best representation of the underlying physics.  

In this case, the lower-lying state is now addressing the Euclidean
time tail of the projected correlator which spoiled the $\chi^2/{\rm
  dof}$ in the fit from $t=20$ to 27 in
Fig.~\ref{fig:meff_n-_350cfg_1600cfg_state2.1By2lngtBygtPlus2}.  The
two-state fit $\chi^2/{\rm dof}$ of 0.10 to 0.16 argues against
dropping any further time slices from the fit.  These small values for
the $\chi^2/{\rm dof}$ are associated with the introduction of not one
but two additional parameters to the fit function.  One needs both an
additional mass and a measure of the relative strengths of the
couplings of the two states to the interpolator.  The presence of
three parameters in the fit function enables an excellent description
of the data that would withstand a significant increase in the
statistical accuracy of the results.

The lower-lying state is suppressed by one to two orders of magnitude
relative to the dominant state in the range $20 \le t \le 23$ included
in the fit of Fig.~\ref{fig:meff_n-_350cfg}.  Here the ratio of
amplitudes $\lambda_1/\lambda_2$ is of order $1/100$ with the
low-lying scattering state making a very small contribution revealed
only through ample Euclidean time evolution.  The range of the
low-lying mass, $M_1$, readily encompasses the infinite volume
scattering threshold of $M_{N} + m_{\pi} = 1.35$ GeV and the
preference for lower lying values is in accord with the anticipated
attractive interaction on the finite volume lattice.

\begin{figure}[t!]
  \begin{center} 
  \includegraphics[height=0.46\textwidth,angle=90]{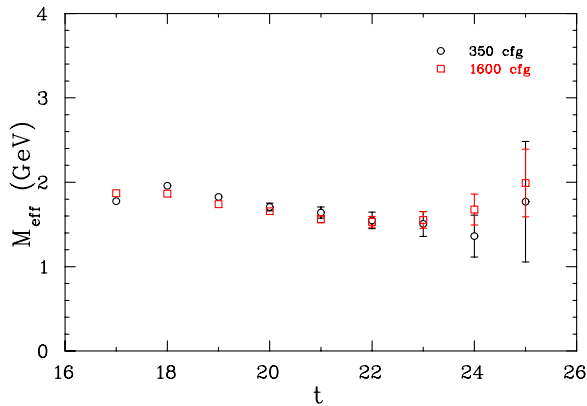}
 \end{center}
\vspace{-0.5cm}
 \caption{(Color online). Effective mass function $M_{\rm eff}(t) =
   (1/2) \log( G(t)/G(t+2))$ for the lowest $N{1/2}^{-}$ state for
   1600 configurations is compared with previous results from 350
   configurations~\cite{Mahbub:2012ri}.  The projected correlation
   functions are obtained from an $8\times 8$ correlation matrix
   analysis at $t_0 = 18$ and $\triangle t = 2$.  As there is no
   evidence for a low-lying scattering state contribution in the
   effective mass tail, the delay of plateau onset is associated with
   excited state contributions. }
 \label{fig:meff_n-_350cfg_1600cfg_state1.1By2lngtBygtPlus2}
\end{figure}

Further support for the more cautious $t_0=19$ analysis illustrated in
the the right hand-panel of Fig.~\ref{fig:data_and_fit} is provided in
Fig.~\ref{fig:meff_n-_350cfg_1600cfg_state1.1By2lngtBygtPlus2}
presenting the effective mass function for the lowest-lying first
odd-parity state observed in our $8\times 8$ correlation matrix
analysis.  In this case there is no evidence for a low-lying
scattering state.
A fit from $t=20$ to 24 inclusive provides a $\chi^2/{\rm dof} = 1.24$
and leaves a 30\% chance of finding a higher $\chi^2/{\rm dof}$ in a
subsequent simulation.

However, significant evolution of the effective mass is observed at
early Euclidean time and it is clear that the projected correlation
function has small admixtures of additional states.  This is due to
more than 8 states participating in the correlation functions of the
correlation matrix and these may include multi-particle scattering
states higher in energy.  Indeed Ref.~\cite{Dudek:2012xn} analysing
four-quark $\pi \pi$ contributions to the isovector vector correlator
of the $\rho$ meson found many $\pi \pi$ scattering contributions
before the first excited state observed when using two-quark
interpolating fields alone.

Given the direct observation herein of a low-lying multi-particle
scattering-state threshold in a ``projected'' correlation function one
must expect similar contributions from the next two-particle
zero-momentum scattering-states having the back-to-back momenta
allowed on the lattice.  Similarly, given the observation of several
two-particle scattering-state contributions in the $\rho$ meson
channel \cite{Dudek:2012xn}, the curvature observed at early times can
be attributed to the higher-energy scattering-state contributions.
Thus the analysis with $t_0 = 19$ in the right-hand plot of
Fig.~\ref{fig:data_and_fit} is the correct representation of the
multi-particle contributions to the nucleon correlator under
examination herein.  Careful consideration of the
$\chi^{2}/{\rm{dof}}$ allows us to circumvent contamination from
higher=lying states and ensure we are extracting the finite-volume QCD
eigenstate energy.

\section{Conclusions}

We have revealed the manner in which the absence of a strong coupling
to multi-particle components of QCD eigenstates can allow scattering
states to be superposed with the dominant state in a projected
correlation function from a correlation matrix analysis.  Even if the
interpolating fields are poor at creating these multi-particle
components, QCD dynamics will ensure their formation in the resolution
of the eigenstates of QCD,

We have explored two interpretations of how states are superposed to
give rise to the observed projected correlation function and
illustrated with reference to a real-world example how this
superposition of states can impact the results extracted from lattice
correlation functions.  Given the direct observation herein of a
low-lying multi-particle scattering-state threshold in a ``projected''
correlation function one must expect similar contributions from the
next two-particle zero-momentum scattering-states having the
back-to-back momenta allowed on the lattice \cite{Dudek:2012xn}.
These states will give rise to curvature in the effective mass at
early Euclidean times and therefore the analysis with $t_0 = 19$
relative to the source at $t_s=16$ in the right-hand plot of
Fig.~\ref{fig:data_and_fit} is the correct representation of the
low-lying multi-particle contribution to the nucleon correlator under
examination herein.

We have discovered that the low-lying scattering states not
observed in Ref.~\cite{Mahbub:2012ri} are hidden within the projected
correlation functions as very small contributions to the correlation
functions suppressed by a factor the order of $1/100$.
In the realm where previous fits were performed, their contribution to
the correlation function is suppressed by one to two orders of
magnitude, as illustrated in the right-hand plot of
Fig.~\ref{fig:data_and_fit}.  As a result, the undetected presence of
a lower-lying scattering state has only a small effect on the
extracted mass.
It is the judicious treatment of the $\chi^{2}/{\rm{dof}}$ that
assists in avoiding systematic errors.

The extent to which one can separate multiple states in a single
correlator has also been illustrated.  It is readily apparent that 
multi-hadron states must be isolated in the correlation matrix
analysis if one is to learn their properties.  While effective
techniques exist to avoid their effects, discovering their properties
is a different matter.

Research is already well underway in exploring the best manner to do
this
\cite{Morningstar:2011ka,Lang:2012db,Kiratidis:2012mr,Morningstar:2013bda}.
The aim is to create correlation matrices composed from three- and
five-quark operators.  Strong coupling to the multi-particle
components of the QCD eigenstates, $|\, 2 \rangle$, is often obtained
by projecting the momentum of each of the hadrons participating in the
scattering state.  Alternative approaches allow the five-quark
operators to have strong overlap with both single-particle dominated
and multi-particle dominated states and alter this overlap through
variation of the fermion propagator source and sink smearing
\cite{Kiratidis:2012mr}.  Through consideration of a variety of
approaches on the same underlying set of gauge field configurations
one can determine the merits of the various approaches and determine
the finite-volume spectrum of QCD in an accurate manner.

\section*{Acknowledgments}
We thank PACS-CS Collaboration for making these $2+1$ flavor
configurations available and the ongoing support of the ILDG.  This
research was undertaken with the assistance of resources at the NCI
National Facility in Canberra, Australia, and the iVEC facilities at
Murdoch University (iVEC@Murdoch) and the University of Western
Australia (iVEC@UWA). These resources were provided through the
National Computational Merit Allocation Scheme, supported by the
Australian Government and the University of Adelaide Partner Share.
We also acknowledge eResearch SA for their supercomputing support
which has enabled this project.  This research is supported by the
Australian Research Council.




\end{document}